\documentclass[aps,prl,twocolumn,showpacs,superscriptaddress]{revtex4-1}

\usepackage{graphicx}
\usepackage[utf8]{inputenc}
\usepackage{caption}
\usepackage{siunitx}
\usepackage{amsmath}
\usepackage{gensymb}
\usepackage{amssymb}
\usepackage{enumerate}
\usepackage{color}
\usepackage{hyperref}
\usepackage[american]{babel}
\usepackage{subcaption}
\usepackage{etoolbox}
\usepackage{floatrow}

\patchcmd{\thebibliography}{\section*{\refname}}{}{}{}

\newcommand{\fig}[1] {Fig.~\ref{fig:#1}}
\newcommand{\tab}[1] {Tab.~\ref{tab:#1}}
\newcommand{\eqn}[1] {Eq.~\ref{eqn:#1}}

\newcommand{\aegis}{AE$\bar{\hbox{g}}$IS}

\newcommand{\pos}{$\mathrm{e^+}$}

\newcommand{\macor}[0]{$\text{MACOR}$}
\newcommand{\off} [0]{{of\!f}}
\newcommand{\on}  [0]{{on}}

\newcommand{\Area}[0]{\mathcal{A}}

\newcommand{\orb}[3]{$#1^{#2}\text{#3}$}
\newcommand{\ttP}[0]{\orb{3}{3}{P}}

\newcommand{\ooS}[0]{\orb{1}{1}{S}}

\newcommand{\otS}[0]{\orb{1}{3}{S}}

\newcommand{\tS}[0]{\orb{2}{3}{S}}
\newcommand{\tP}[0]{\orb{2}{3}{P}}

\DeclareSIUnit\inch{in.}
\DeclareSIUnit\division{div}

\begin{document}

\preprint{APS/123-QED}

\title{Velocity selected production of $ 2^3\text{S} $ metastable positronium}

\newcommand{\corresponding}[1]{\altaffiliation{Corresponding author, #1}}

\newcommand{\affpolimi}[0]{\affiliation{LNESS, Department of Physics, Politecnico di Milano, via Anzani 42, 22100~Como, Italy}}
\newcommand{\affinfnmi}[0]{\affiliation{INFN, Sezione di Milano, via Celoria 16, 20133~Milano, Italy}}
\newcommand{\affvienna}[0]{\affiliation{Stefan Meyer Institute for Subatomic Physics, Austrian Academy of Sciences, Boltzmanngasse 3, 1090~Vienna, Austria}}
\newcommand{\affinsubria}[0]{\affiliation{Department of Science and High Technology, University of Insubria, Via Valleggio 11, 22100~Como, Italy}}
\newcommand{\affjinr}[0]{\affiliation{Joint Institute for Nuclear Research, Dubna~141980, Russia}}
\newcommand{\affbs}[0]{\affiliation{Department of Mechanical and Industrial Engineering, University of Brescia, via Branze 38, 25123~Brescia, Italy}}
\newcommand{\affinfnpv}[0]{\affiliation{INFN Pavia, via Bassi 6, 27100~Pavia, Italy}}
\newcommand{\afftn}[0]{\affiliation{Department of Physics, University of Trento, via Sommarive 14, 38123~Povo, Trento, Italy}}
\newcommand{\affinfntn}[0]{\affiliation{TIFPA/INFN Trento, via Sommarive 14, 38123~Povo, Trento, Italy}}
\newcommand{\affge}[0]{\affiliation{Department of Physics, University of Genova, via Dodecaneso 33, 16146~Genova, Italy}}
\newcommand{\affinfnge}[0]{\affiliation{INFN Genova, via Dodecaneso 33, 16146~Genova, Italy}}
\newcommand{\affmi}[0]{\affiliation{Department of Physics ``Aldo Pontremoli'', Universit\`{a} degli Studi di Milano, via Celoria 16, 20133~Milano, Italy}}
\newcommand{\affmpi}[0]{\affiliation{Max Planck Institute for Nuclear Physics, Saupfercheckweg 1, 69117~Heidelberg, Germany}}
\newcommand{\afflac}[0]{\affiliation{Laboratoire Aim\'e Cotton, Universit\'e Paris-Sud, ENS Paris Saclay, CNRS, Universit\'e Paris-Saclay, 91405~Orsay Cedex, France}}
\newcommand{\affpolimiII}[0]{\affiliation{Department of Aerospace Science and Technology, Politecnico di Milano, via La Masa 34, 20156~Milano, Italy}}
\newcommand{\affheidelberg}[0]{\affiliation{Kirchhoff-Institute for Physics, Heidelberg University, Im Neuenheimer Feld 227, 69120~Heidelberg, Germany}}
\newcommand{\affcern}[0]{\affiliation{Physics Department, CERN, 1211~Geneva~23, Switzerland}}
\newcommand{\affoslo}[0]{\affiliation{Department of Physics, University of Oslo, Sem Saelandsvei 24, 0371~Oslo, Norway}}
\newcommand{\afflyon}[0]{\affiliation{Institute of Nuclear Physics, CNRS/IN2p3, University of Lyon 1, 69622~Villeurbanne, France}}
\newcommand{\affmoscow}[0]{\affiliation{Institute for Nuclear Research of the Russian Academy of Science, Moscow~117312, Russia}}
\newcommand{\affinfnpd}[0]{\affiliation{INFN Padova, via Marzolo 8, 35131~Padova, Italy}}
\newcommand{\affprague}[0]{\affiliation{Czech Technical University, Prague, Brehov\'a 7, 11519~Prague~1, Czech Republic}}
\newcommand{\affbo}[0]{\affiliation{University of Bologna, Viale Berti Pichat 6/2, 40126~Bologna, Italy}}
\newcommand{\affpv}[0]{\affiliation{Department of Physics, University of Pavia, via Bassi 6, 27100~Pavia, Italy}}
\newcommand{\affnorway}[0]{\affiliation{The Research Council of Norway, P.O. Box 564, 1327~Lysaker, Norway}}
\newcommand{\affheidelbergII}[0]{\affiliation{Department of Physics, Heidelberg University, Im Neuenheimer Feld 226, 69120~Heidelberg, Germany}}
\newcommand{\affbsII}[0]{\affiliation{Department of Civil, Environmental, Architectural Engineering and Mathematics, University of Brescia, via Branze 43, 25123~Brescia, Italy}}


\author{C.~Amsler}
\affvienna

\author{M.~Antonello}
\affinfnmi
\affinsubria

\author{A.~Belov}
\affmoscow

\author{G.~Bonomi}
\affbs
\affinfnpv

\author{R.~S.~Brusa}
\afftn
\affinfntn

\author{M.~Caccia}
\affinfnmi
\affinsubria

\author{A.~Camper}
\affcern

\author{R.~Caravita}
\corresponding{ruggero.caravita@cern.ch}
\affcern

\author{F.~Castelli}
\affinfnmi
\affmi

\author{G.~Cerchiari}
\affmpi

%

\author{D.~Comparat}
\afflac

\author{G.~Consolati}
\affpolimiII
\affinfnmi

\author{A.~Demetrio}
\affheidelberg

\author{L.~Di~Noto}
\affge
\affinfnge

\author{M.~Doser}
\affcern


\author{M.~Fan\`{i}}
\affge
\affinfnge
\affcern




\author{S.~Gerber}
\affcern


\author{A.~Gligorova}
\affvienna

\author{F.~Guatieri}
\afftn
\affinfntn

\author{P.~Hackstock}
\affvienna

\author{S.~Haider}
\affcern

\author{A.~Hinterberger}
\affcern

\author{H.~Holmestad}
\affoslo

\author{A.~Kellerbauer}
\affmpi

\author{O.~Khalidova}
\affcern

\author{D.~Krasnick\'y}
\affinfnge

\author{V.~Lagomarsino}
\affge
\affinfnge

\author{P.~Lansonneur}
\afflyon

\author{P.~Lebrun}
\afflyon

\author{C.~Malbrunot}
\affcern
\affvienna

\author{S.~Mariazzi}
\corresponding{mariazzi@science.unitn.it}
\afftn
\affinfntn


\author{V.~Matveev}
\affmoscow
\affjinr


\author{S.~R.~M\"{u}ller}
\affheidelberg

\author{G.~Nebbia}
\affinfnpd

\author{P.~Nedelec}
\afflyon

\author{M.~Oberthaler}
\affheidelberg

\author{D.~Pagano}
\affbs
\affinfnpv

\author{L.~Penasa}
\afftn
\affinfntn

\author{V.~Petracek}
\affprague

\author{F.~Prelz}
\affinfnmi

\author{M.~Prevedelli}
\affbo

\author{B.~Rienaecker}
\affcern

\author{J.~Robert}
\afflac

\author{O.~M.~R{\o}hne}
\affoslo

\author{A.~Rotondi}
\affinfnpv
\affpv

\author{H.~Sandaker}
\affoslo

\author{R.~Santoro}
\affinfnmi
\affinsubria

\author{L.~Smestad}
\affcern
\affnorway

\author{F.~Sorrentino}
\affge
\affinfnge

\author{G.~Testera}
\affinfnge

\author{I.~C.~Tietje}
\affcern


\author{E.~Widmann}
\affvienna

\author{P.~Yzombard}
\affmpi

\author{C.~Zimmer}
\affcern
\affmpi
\affheidelbergII


\author{N.~Zurlo}
\affinfnpv
\affbsII

\collaboration{The AEgIS collaboration}
\noaffiliation{}


\date{\today}


\begin{abstract}
Positronium in the \tS{} metastable state exhibits a low electrical polarizability and a long lifetime (\SI{1140}{\nano\second}) making it a promising candidate for interferometry experiments with a neutral matter-antimatter system. In the present work, \tS{} positronium is produced - in absence of electric field - via spontaneous radiative decay from the \ttP{} level populated with a \SI{205}{\nano\meter} UV laser pulse. Thanks to the short temporal length of the pulse, \SI{1.5}{\nano\second} full-width at half maximum, different velocity populations of a positronium cloud emitted from a nanochannelled positron/positronium converter were selected by delaying the excitation pulse with respect to the production instant. \tS{} positronium atoms with velocity tuned between $ 7 \cdot 10^4 \, \si{\meter\per\second} $ and $ 10 \cdot 10^4 \, \si{\meter\per\second} $ were thus produced. Depending on the selected velocity, a \tS{} production efficiency ranging from $ \sim 0.8 \% $ to $ \sim 1.7 \% $, with respect to the total amount of emitted positronium, was obtained. The observed results give a branching ratio for the \ttP{}--\tS{} spontaneous decay of $ (9.7 \pm 2.7) \% $. The present velocity selection technique could allow to produce an almost monochromatic beam of $ \sim 1 \cdot 10^3 $ \tS{} atoms with a velocity spread $ < 10^4 \si{\meter\per\second} $ and an angular divergence of $ \sim \SI{50}{\milli\radian} $.
\end{abstract}

\maketitle{}

Positronium (Ps) is one of the few matter/antimatter systems (with antihydrogen and muonium) being considered for probing experimentally the gravitational interaction \cite{cassidy_review:18, alpha_grav:13, aegis_natc:14, gbar:15, mage:18}. 
Several experimental schemes based on long-lived Ps beams have been proposed, either letting the atoms free-fall in a drift tube \cite{mills_leventhal:02, cassidy_psgravity:14} or by using a matter-wave atom interferometer  \cite{oberthaler_ps:02} to measure their vertical displacement with a position-sensitive detector. Gravity (or any other force acting on the atoms) can then be worked out if the average velocity of the atoms is known \cite{aegis_natc:14, oberthaler_ps:02}.

All the suggested schemes involve laser excitation to long-lived excited states to overcome the lifetime limitation of the \otS{} ground state (\SI{142}{\nano\second}) where Ps is normally produced. An effective and widely adopted choice consists in laser-exciting the atoms to the Rydberg levels, where lifetimes spanning from tens of \si{\micro\second} up to several \si{\milli\second} can be obtained \cite{cassidy_ryd:12, cassidy_ryd:15}. 
Atoms in Rydberg states are, in general, sensitive to electric field gradients \cite{hogan_rydrev:16} which can modify their trajectories. This is due to their large electrical polarizability (up to $ \sim 10^{-32} \, \si{\coulomb\meter\squared\per\volt} $ for Ps in $ n = 15 $) \cite{jones_prl:17, alonso_guide:17}. Ps Rydberg sublevels with large electrical dipole can be guided and focused \cite{jones_prl:17,alonso_guide:17} while the selective excitation to sublevels with low dipole moment has been proposed \cite{jones_ryd:16} as a method to minimize the deflection of Rydberg Ps in interferometric measurements with physical gratings \cite{alonso_metastable:17}.


An alternative way to produce a beam of long-lived Ps with lower electrical polarizability ($ \sim  10^{-38} \,\, \si{\coulomb\meter\squared\per\volt} $) consists in laser exciting the atoms to their \tS{} metastable level \cite{alonso_metastable:17}, whose lifetime is \SI{1.14}{\micro\second} in vacuum and in the absence of electric field \cite{aegis_meta:18}. A beam of \tS{} Ps atoms (of known average velocity) has been shown to be suitable for improving the inertial sensitivity in proposed matter-wave interferometric layouts \cite{oberthaler_ps:02}. Moreover, the availability of \tS{} Ps with average velocity $ < 10^5 \, \si{\meter\per\second} $ would allow keeping the interferometer compact in length ($ L \lesssim \si{1\thinspace  \meter} $), thus easing the control of thermal and vibrational noise \cite{miffre_vibrnoise:06}.

Producing fast \tS{} Ps with energies of several \si{\electronvolt} has already been demonstrated via \pos{} collisions with solid \cite{canter:75, steiger:92, day:01} or gaseous targets \cite{laricchia:85, murtagh:09}. \tS{} Ps atoms with a beam Maxwellian distribution at around 600 K (average speed $ > 1.4 \cdot 10^5 \, \si{\meter\per\second} $) were also produced via Doppler-free two photon excitation of ground-state atoms of Ps desorbed from metallic surfaces \cite{chu_mills:82, fee_1s2s_prl:93, fee_1s2s_pra:93} and \otS{}--\tS{} two photon excitation of Ps emitted from porous silicon \cite{cooke_crivelli:15}. Production of \tS{} Ps via single photon excitation of ground-state atoms to mixed \tS{}--\tP{} in electric fields \cite{alonso_metastable:17} and the single photon excitation of ground-state atoms to \ttP{} levels with subsequent radiative decay to \tS{} \cite{aegis_meta:18} have been recently demonstrated. The reduction of the \tS{} lifetime in electric fields due to Stark mixing has also been studied \cite{aegis_meta:18}.

In the present work we investigate the feasibility of a source of metastable \tS{} Ps with defined and tunable velocity in the absence of electric field. The \tS{} level is populated by spontaneous radiative decay of laser-excited \ttP{} Ps atoms. The tuning of the \tS{} Ps velocity is achieved by varying the delay of the \otS{}--\ttP{} excitation pulse between \SI{20}{\nano\second} and \SI{65}{\nano\second} from the \pos{} implantation time in a nanochannelled silicon \pos{}--Ps converter \cite{mariazzi_prb:10}, thus selecting Ps populations emitted after different permanence times in the target \cite{cassidy_perman:10, mariazzi_jphys:15} and consequently with different velocities \cite{mariazzi_prl:10, phd_guatieri:18}.

\begin{figure}[htp]
\centering
\includegraphics[width=\linewidth]{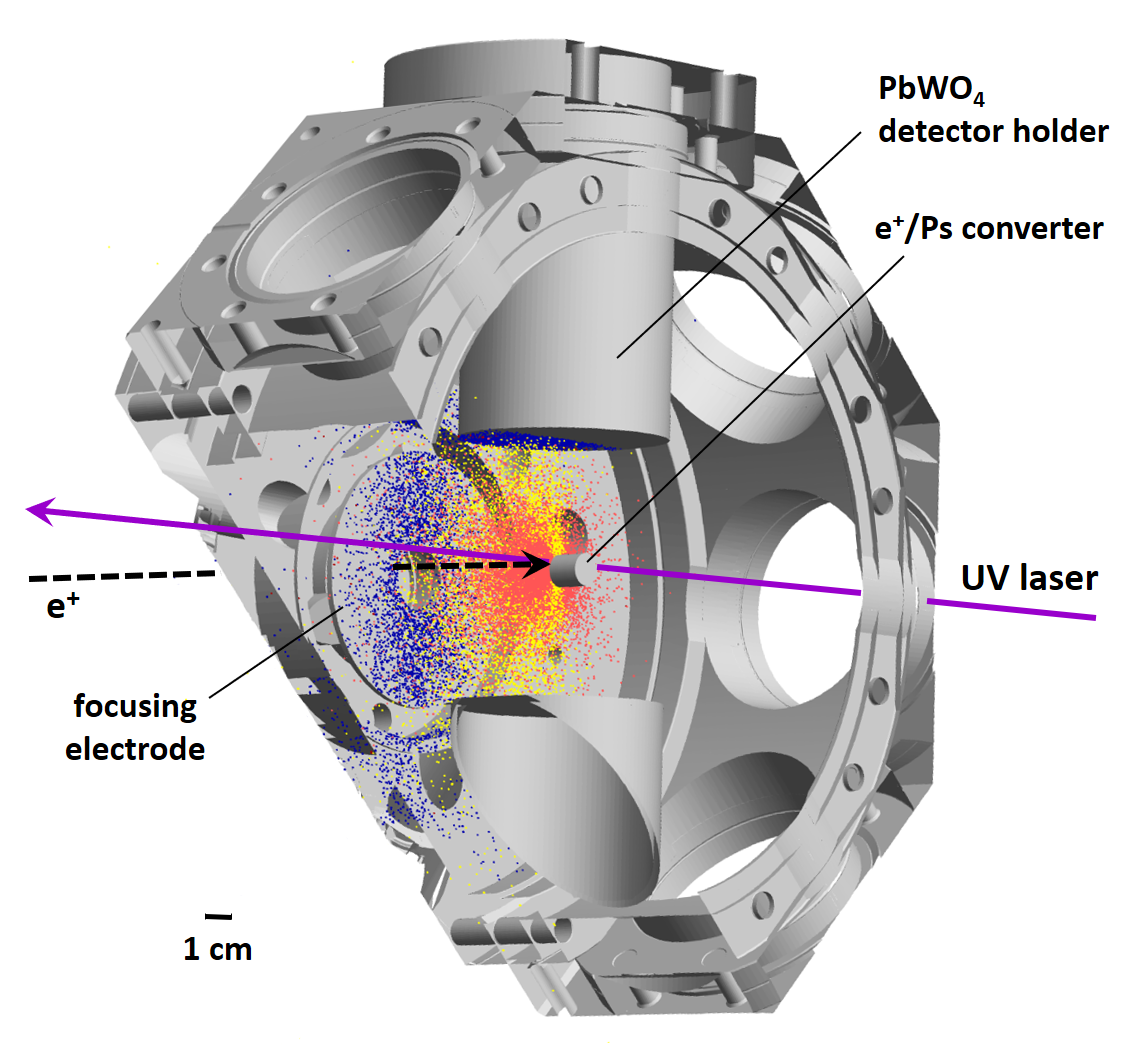}
\caption{
Schematic of the experimental chamber with example Monte Carlo in-flight and collision annihilation distributions for an \otS{} Ps population (light and dark red dots) and for a \tS{} Ps population (yellow and blue dots). The line-shaped annihilations' distribution of \tS{} atoms is due to the UV laser Doppler selection.}
\label{fig:mc}
\end{figure}

In our experiment, Ps is formed when $ \sim \SI{7}{\nano\second} $ bursts of $ \sim 10^7 $ \pos{}, prepared in the \aegis{} \pos{} system (see \cite{aegis_meta:18, aegis_nimb:15, aegis_neq3:16} for a detailed description of the apparatus) are electro-magnetically transported and implanted at \SI{3.3}{\kilo\electronvolt} into a Si (111) p-type crystal where nanochannels were previously produced via electrochemical etching and thermal oxidized in air \cite{aegis_meta:18, aegis_neq3:16}. Ps produced inside the converter out-diffuses into vacuum through the nanochannels loosing a fraction of its emission energy by collision with the walls.  The Si target was kept at room temperature and \pos{} were implanted into it with a spot of $ \sim \SI{3}{\milli\meter} $ in size. Measurements previously performed on identical \pos{}--Ps converters indicated a wide angular emission of Ps from the nanochannels \cite{aegis_meta:18, aegis_neq3:16}. A schematic of the experimental chamber is illustrated in \fig{mc}. 

In the target region, \pos{} are guided by a \SI{25}{\milli\tesla} magnetic field and focused by an electrostatic lens formed by the last electrode of the transfer line set at \SI{-3000}{\volt} and the target kept at ground potential (see \fig{mc} for the geometry), inducing an electric field of about \SI{300}{\volt\per\centi\meter} in front of the converter. Since this electric field shortens considerably the \tS{} lifetime \cite{alonso_guide:17, aegis_meta:18}, for the present measurements the focusing electrode was switched off $ \sim \, \SI{5}{\nano\second} $ after the \pos{} implantation using a fast switch with a rise time of $ \sim \SI{15}{\nano\second} $ (from \SI{-3000}{\volt} to \SI{0}{\volt}). 

Ps emitted into vacuum was subsequently excited from the \otS{} ground state to \ttP{} sublevel manifold with an UV laser pulse set at the transition resonance wavelength $ \lambda_{UV} = (205.045 \pm 0.005) \, \si{\nano\meter} $. The laser setup is described in detail elsewhere \cite{aegis_neq3:16, master_caravita:13, cialdi_nimb:11}. The UV pulse energy was kept above \SI{60}{\micro\joule} and the effective size of the spot was about $ 3.0-3.5 \, \si{\milli\meter} $ full width at half maximum (FWHM) both in horizontal and vertical direction in front of the target. The laser beam was aligned grazing the target and its position and size were monitored with a CCD camera on a \SI{1}{\inch} \macor{} screen placed inside the vacuum region (a few \si{\centi\meter} away from the target).  The UV pulse had a horizontal polarization (i.e. perpendicular to the target), a nearly Gaussian temporal profile with a FWHM of \SI{1.5}{\nano\second}, and a Gaussian-like spectral profile with a bandwidth $ \sigma_{UV} = 2\pi \,\times\, \SI{120}{\giga\hertz} $. 


The laser bandwidth is narrower than the Doppler profile of Ps emitted from the used \pos{}--Ps converter ($ \sim 2\pi \,\times\, \SI{470}{\giga\hertz} $, see \cite{aegis_neq3:16}). As a consequence, the UV pulse selectively excites to \ttP{} only the fraction of the emitted atoms with a velocity component parallel to the laser propagation axis, $v_\parallel$, with $|v_\parallel|< 2.5 \cdot 10^4 \, \si{\meter\per\second} $. Following the excitation, a fraction of Ps in the \ttP{} sublevel spontaneously decays to \tS{} emitting a \SI{1312}{\nano\meter} photon with an expected branching ratio of 10 \% in the presence of a \SI{25}{\milli\tesla} magnetic field \cite{aegis_meta:18}. The atoms in the \tS{} sublevel retain - with a good approximation - the original velocity distribution, since both recoil velocities for the absorption of a \SI{205}{\nano\meter} photon ($ \approx 1.8 \cdot 10^3 \, \si{\meter\per\second} $) and the emission of a \SI{1312}{\nano\meter} photon ($ \approx 2.8 \cdot 10^2 \, \si{\meter\per\second} $) are negligible. The UV pulse was delayed, with respect to the \pos{} implantation time, from \SI{20}{\nano\second} (to let the electric field reach \SI{0}{\volt\per\centi\meter}) up to \SI{65}{\nano\second} by using a SRS DG645 digital delay generator.

\begin{figure}[htp]
\centering
\includegraphics[width=\linewidth, clip]{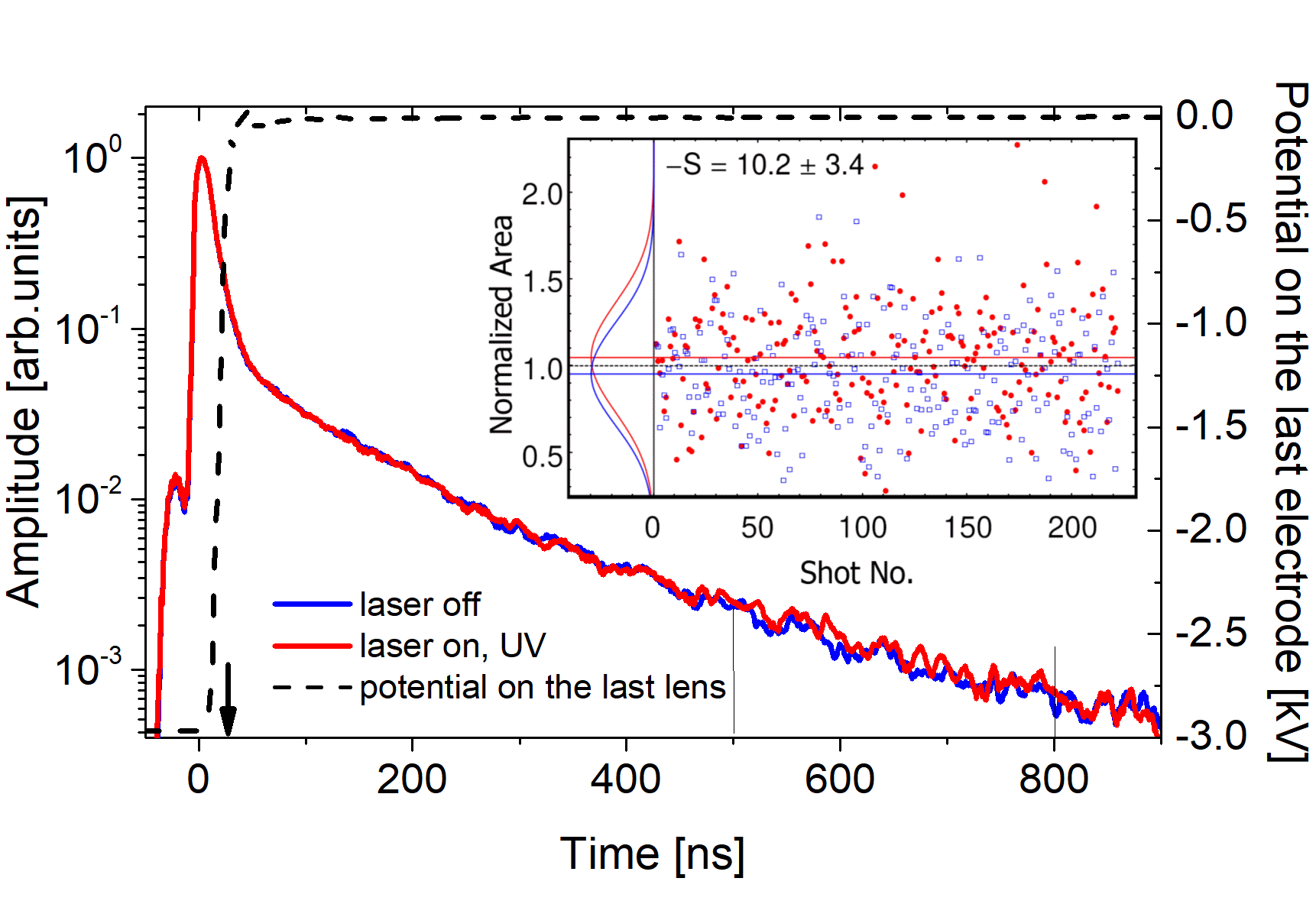}
\caption{Average of 200 SSPALS spectra of Ps in vacuum with UV laser off (blue) and on (red), normalized to the peak height. The potential on the focusing electrode is also shown (dashed line).
The arrow marks the laser pulse instant (here \SI{20}{\nano\second} after the peak) and the vertical lines delimit the area used to conduct the detrending analysis of the $S$ parameter shown in the inset (see text and \cite{aegis_meta:18}). Full circles and empty squares are the detrended areas with/ without laser (i.e. $ {\Area_\on}^i $ and  $ {\Area_\off}^i $). The sideways Gaussian curves are the distributions of $ {\Area_\on}^i $ (red) and $ {\Area_\off}^i $ (blue).}
\label{fig:sspals_uvonly}
\end{figure}

The time distribution of gamma rays emitted by \pos{} and Ps annihilations, i.e. the so-called Single-Shot Positron Annihilation Lifetime Spectroscopy (SSPALS) spectrum, was acquired with the same procedure used in \cite{aegis_meta:18, aegis_neq3:16}. A $ 20 \times 25 \times 25$-$\, \si{\milli\meter} $ $ \mathrm{{PbWO}_4} $ scintillator, coupled to a Hamamatsu R11265-100 photomultiplier tube (PMT) and digitized by a HD4096 Teledyne LeCroy oscilloscope, was placed \SI{40}{\milli\meter} above the target (\fig{mc}). In the presence of Ps formation, SSPALS spectra present a prompt peak, given by the fast $ 2\gamma $ annihilations of \pos{} implanted in the target, and a tail that is dominated by the $ 3\gamma $ decay of Ps emitted into vacuum (\fig{sspals_uvonly}). The changes in the Ps population induced by the interaction with laser light affect the area under the tail. This effect can be quantified by using the $S$ parameter evaluated as: 


\begin{equation}%
	S ~=~ \frac{\Area_\off - \Area_\on}{\Area_\off} \, ,
	\label{eqn:sdef}
\end{equation}

\noindent
where $ \Area_\off $ and $ \Area_\on $ are the averages of the normalized areas $ {\Area_\off}^i $ and $ {\Area_\on}^i $ below the $i$-th SSPALS shot, calculated in a given time window with lasers off and on respectively \cite{aegis_neq3:16}. The areas were normalized by using the detrending procedure described in \cite{aegis_meta:18}, which mitigates the effect of eventual slow drifts in the positron beam intensity. The fraction of Ps excited to \ttP{} with the UV pulse was evaluated by selectively photoionizing the excited atoms with an IR pulse ($ \lambda_{IR} = \SI{1064}{\nano\meter} $, energy of \SI{50}{\milli\joule} and temporal FWHM of \SI{4}{\nano\second} \cite{aegis_neq3:16}). The \otS{}--\ttP{} photoionization process results in a decrease of the Ps population decaying into $ 3\gamma $ at late times. 


Samples of 200 single SSPALS shots alternating UV+IR lasers on and off were collected with different laser delays to measure the \otS{}--\ttP{} excitation efficiency. A $ S = (13.8 \,\pm\, 2.2) \, \% $ was observed in the selected $ 50-500 \, \si{\nano\second} $ time window after the prompt peak for a laser delay of \SI{20}{\nano\second}, in agreement with the results of Ref. \cite{aegis_neq3:16, aegis_meta:18}. Repeating the measurement delaying the UV+IR laser pulses to \SI{35}{\nano\second} and \SI{50}{\nano\second}, the $ S $ signal decreased by $ \sim 30\% $, to $ S = (8.8 \pm 2.6) \% $, and by $ \sim 60\% $, to $ S = (6.8 \pm 2.9) \% $, respectively, since an increasing fraction of Ps atoms already left the target proximity before the laser pulse. 

Sending the UV laser only, a fraction of the excited Ps is allowed to spontaneously decay from \ttP{} to \tS{}. In absence of an electric field, as in the present case, the lifetime of Ps in the \tS{} state is eight times longer than in the \otS{} state (\SI{1140}{\nano\second} vs \SI{142}{\nano\second}) \cite{cassidy_review:18, aegis_meta:18}. Due to this longer lifetime, a larger fraction of atoms survives in-flight annihilation and reaches the experimental chamber walls, where it annihilates in $ 2\gamma$ with an electron of the medium producing a signal excess clearly identifiable in SSPALS spectra (\fig{sspals_uvonly}).

A previously developed Monte Carlo (MC) code \cite{aegis_meta:18} was used to calculate the expected spatial and the temporal distributions of \otS{} and \tS{} annihilations in the geometry of our experimental chamber. This code calculates the atoms' flight trajectories and position and time annihilation distributions, accounting for both in-flight self-annihilations and collisions with the walls, by integrating the equation of motion for the single particles and the optical rate equations of the internal level dynamics, while testing for collisions with the chamber walls. The final annihilation distributions are obtained upon averaging over the sample of atoms sorted on the initial velocity and position distributions. The free parameters are the \otS{}--\ttP{} excitation efficiency (obtained experimentally from the UV+IR data), the \ttP{}--\ooS{} quenching efficiency (assumed to be 17\% in our \SI{25}{\milli\tesla} magnetic field \cite{aegis_neq3:16, aegis_meta:18}), the \ttP{}--\tS{} branching efficiency (to be determined), plus those specifying the initial position/velocity distributions.
As no external electric field is present and the expected maximum motional Stark electric field is $ \approx \SI{25}{\volt\per\centi\meter} $ in our \SI{25}{\milli\tesla} magnetic field with a Ps velocity of the order of $ \approx 10^5 \, \si{\meter\per\second} $ (corresponding to an optical decay rate of \tS{} in electric field of $ \approx \SI{100}{\kilo\hertz} $ \cite{aegis_meta:18}), the optical decay of \tS{} atoms was not included in the simulations. Ultimately the code produces a model estimate of the $ S $ parameter that can be directly compared with experimental data. 

In order to give a pictorial representation of the expected annihilation positions of the \otS{} and \tS{} populations, a MC simulation has been conducted with $ 10^4 $ Ps atoms in each population (i.e. not reflecting the real \tS{}/\otS{} relative abundance to emphasize the rare annihilations of the \otS{} fraction on the walls) and setting an uniform velocity of $ 10^5 \, \si{\meter\per\second} $ in modulus for all Ps atoms (\fig{mc}). Isotropic emission of \otS{} Ps from the converter and the Doppler selection from a $ 2\pi \,\times\, \SI{120}{\giga\hertz} $ excitation laser were assumed \cite{aegis_neq3:16, phd_caravita:17, phd_guatieri:18}. According to the MC simulations, around 60\% of the produced \tS{} Ps reaches the walls as opposed to only $ \sim 1\% $ of \otS{} atoms with the same velocity. This causes the additional annihilations in the tail of the SSPALS spectra shown in \fig{sspals_uvonly}, leading to negative values of the $ S $ parameter (\eqn{sdef}).


Experimentally, when the UV laser delay is set to \SI{20}{\nano\second}, the excess of annihilations due to \tS{} atoms reaching the chamber walls is observed between \SI{500}{\nano\second} and \SI{800}{\nano\second} from the peak maximum, giving a $ S \,=\, (-10.2 \,\pm\, 3.4) \, \% $ (\fig{sspals_uvonly}, inset). The time when this excess appears can be controlled by varying the laser pulse delay.

\begin{table*}[htb]
\begin{tabular*}{0.8 \linewidth}{@{\extracolsep{\fill}}lccc}
      \textbf{Laser delay} & \textbf{\otS{} $ \rightarrow $ \ttP{} efficiency} & \textbf{\ttP{} $ \rightarrow $ \tS{} efficiency} & \textbf{\tS{} average velocity} \\
      \hline
      \SI{20}{\nano\second} & $ (13.8 \pm 2.2) \,\% $ & $ (9.7 \pm 2.7) \,\% $ & $ (1.0 \,\pm\, 0.1) \cdot 10^5 \,\, \si{\meter\per\second} $ \\
      \SI{35}{\nano\second} & $ (8.8 \pm 2.6) \,\% $ & $ (8.7 \pm 5.0) \,\% $ & $ (0.8 \,\pm\, 0.1) \cdot 10^5 \,\, \si{\meter\per\second} $ \\
      \SI{50}{\nano\second} & $ (6.8 \pm 2.9) \,\% $ & $ (10.1 \pm 6.2) \,\% $ & $ (0.7 \,\pm\, 0.1) \cdot 10^5 \,\, \si{\meter\per\second} $ \\          
\end{tabular*}  
\caption{Comparison of the experimental \otS{}--\ttP{} efficiency (from photoionization) and the best found parameters with the Monte Carlo model for different laser delays.} 
\label{tab:results}
\end{table*}

\begin{figure}[htb]
\includegraphics[width=\linewidth, clip]{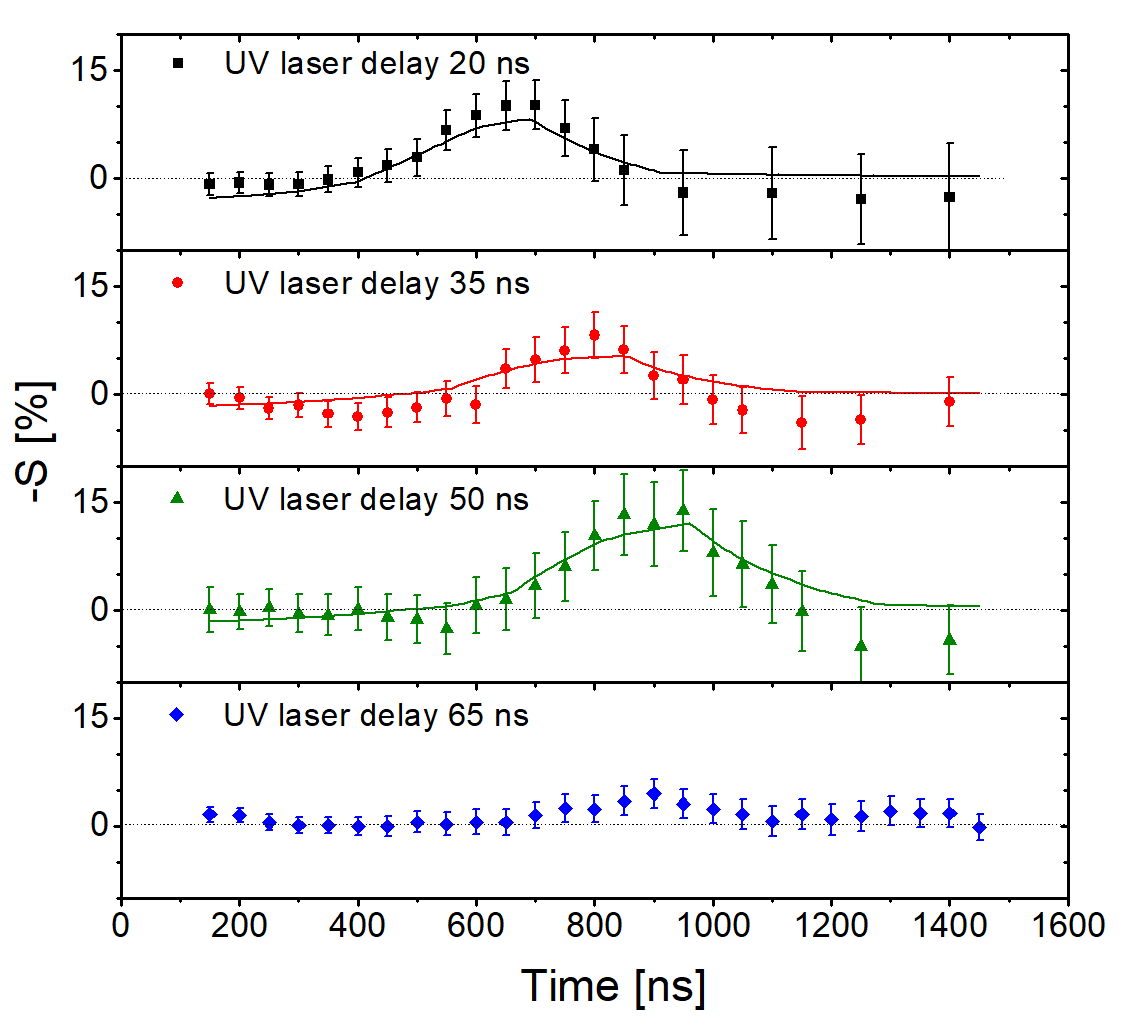}  
\caption{$ -S $ parameter as a function of the time elapsed from the prompt peak for different UV laser delays (20, 35, 50 and 65 \si{\nano\second}). The $ -S $ parameter was calculated using time windows of \SI{300}{\nano\second} in steps of \SI{50}{\nano\second}. The continuous lines are the Monte Carlo best fits (see text) of the $ -S $ parameter. The weak signal at \SI{65}{\nano\second} does not allow to perform a reliable fit.}
\label{fig:time_scan}
\end{figure}

Indeed, nanochanneled Si \pos{}--Ps converters emit Ps in a broad distribution of velocities \cite{mariazzi_prl:10, phd_guatieri:18}. Emission velocity is mostly dependent on the time needed to escape from the nanochannels into vacuum (or permanence time \cite{cassidy_perman:10, mariazzi_jphys:15}): the highest values correspond to the shortest Ps permanence times (\SI{1}{\nano\second} or less) and the lowest ones to the longest permanence times (up to \SI{20}{\nano\second}) \cite{mariazzi_jphys:15, phd_guatieri:18}. As a matter of fact, the fraction that resides longer in the nanochannels loses more energy by collision with the walls. The UV laser pulse, thanks to its brief duration and its limited spatial dimension, excites only the fraction of the emitted Ps in front of the target at the time when the laser is shot. Ps atoms with different permanence times and consequently different velocities can be thus selected varying the laser timing.

Measurements as in \fig{sspals_uvonly} were repeated by retarding the UV laser shot to 35, 50 and 65 \si{\nano\second}. The $ -S $ parameter was evaluated as a function of the time elapsed from the prompt peak in time windows of \SI{300}{\nano\second} with steps of \SI{50}{\nano\second} starting from \SI{150}{\nano\second} to highlight the excesses of Ps annihilations. The curves corresponding to the spectra acquired with the different delay times are reported in \fig{time_scan} (each spectrum is obtained from a sample of 200 single shots).

The plot shows that the time of the annihilations excess, ascribable to \tS{} (i.e. higher values of $-S$), is progressively time shifted by delaying the UV pulse. With a delay of \SI{20}{\nano\second}, the largest fraction of \tS{} atoms reaches the walls of the chamber after 650-700 \si{\nano\second} from the prompt (as also seen in \fig{sspals_uvonly}). This time increases to \SI{800}{\nano\second} and \SI{900}{\nano\second} when the UV pulse is delayed to \SI{35}{\nano\second} and \SI{50}{\nano\second}, respectively. The excess of annihilations almost disappeared with \SI{65}{\nano\second}, indicating that the largest part of the Ps emitted by the converter has already left the laser spot. A rough estimation of the average velocity (for each delay setting) is prompty obtained considering the average distance between the converter and the front wall (\SI{6}{\centi\meter}, see \fig{mc}): $ \SI{6}{\centi\meter} / \SI{650}{\nano\second} \simeq 1.0 \cdot 10^5 \si{\meter\per\second} $, $ \SI{6}{\centi\meter}/\SI{800}{\nano\second} \simeq 8.0 \cdot 10^4 \si{\meter\per\second} $ and $ \SI{6}{\centi\meter}/\SI{900}{\nano\second} \simeq 7.0 \cdot 10^4 \si{\meter\per\second} $ for UV pulse delays of \SI{20}{\nano\second}, \SI{35}{\nano\second} and \SI{50}{\nano\second} respectively. 



A more accurate estimation of the \tS{} velocity distribution was obtained by fitting the experimental \textit{-S} versus time curves (\fig{time_scan}) with the previously introduced MC model, assuming the initial \tS{} Ps atoms velocities to be distributed as a 1D-Gaussian function in modulus and uniformly in angle. The initial position distribution was assumed point-like as the \pos{} spot radius $ \sim \SI{1.5}{\milli\meter} $ is much smaller than other distances at play. 

The superimposed solid lines in \fig{time_scan} were obtained from the MC varying the \ttP{}--\tS{} efficiency and the average velocity in order to find the best agreement between data and predictions. The best-fit parameters and their statistical errors are summarized in \tab{results}. The MC model fit agrees with the previous rough estimation of the average \tS{} velocity for each delay setting. Moreover, it constraints the standard deviation of each \tS{} velocity distribution to $ \Delta v < 1 \cdot 10^4 \, \si{\meter\per\second} $ in all the three cases, pointing out that this source of \tS{} is roughly monochromatic to $ {\Delta v}/v \lesssim 14 \% $ or better (the limit of our current sensitivity). A branching efficiency for the \ttP{}--\tS{} transition of around $ 10 \% $, in agreement with the expected theoretical value \cite{aegis_meta:18}, was also found (\tab{results}). Note that found values and error bars may be affected by unaccounted systematical uncertainties due the approximated time, position and velocity distributions of emitted Ps and laser excitation dynamics. However the results were verified not to be changing significantly (i.e. by more than $ 1.0 \cdot 10^4 \, \si{\meter\per\second} $) using other reasonable Ps emission models, for instance uniform velocity and non-uniform angular emission with an angle-cut at $60^{\circ}$ or higher \cite{phd_caravita:17, phd_guatieri:18}.

In conclusion, we have demonstrated the possibility to produce \tS{} Ps via \ttP{} excitation and spontaneous decay with a selected average velocity in the range 7 - 10 $ \cdot 10^4 \, \si{\meter\per\second} $ and with  $ \Delta v < 1 \cdot 10^4 \, \si{\meter\per\second} $ ($<$ 14\% monochromaticity) in the absence of electric field. 
With present positron-Ps converters \cite{aegis_nimb:15, cassidy_review:18} and bursts of $ 10^7 $ \pos{} \cite{surko_rev:15, liszkay_linac:13}, $ 3-8 \cdot 10^4 $ Ps atoms every minute in \tS{} level can be obtained (according to the selected velocity range). Since the UV laser also acts as angular selector due to its limited bandwidth, these \tS{} Ps atoms expand in the space as depicted in \fig{mc}. Using a proper iris, every minute an almost monochromatic pulsed beam of $ \approx 1 \cdot 10^3 $ \tS{} atoms with an angular divergence of \SI{50}{\milli\radian} could be obtained.
An enhancement of the beam intensity, while retaining the observed velocity selection, could be envisaged using stimulated emission, i.e. increasing the \ttP{}--\tS{} transition rate and its branching efficiency.
Further reduction of the \tS{} Ps atoms velocity looks also feasible thanks to the observed thermal Ps emitted by similar \pos{}--Ps converters when held at cryogenic temperatures \cite{mariazzi_prl:10}. Selecting this thermal fraction should allow to produce monochromatic \tS{} with velocity in the range of low $ 10^4 \, \si{\meter\per\second} $.
The development of long-lived \tS{} Ps beams - with defined and tunable velocity - could open the possibility to perform interferometry measurements with Ps \cite{oberthaler_ps:02}. 

\vspace{0.5cm}

The authors are grateful to Dr. S. Cialdi for the original development of the \otS{}-\ttP{} laser. This work was supported by Istituto Nazionale di Fisica Nucleare; the CERN Fellowship programme and the CERN Doctoral student programme; the Swiss National Science Foundation Ambizione Grant (No. 154833); a Deutsche Forschungsgemeinschaft research grant; an excellence initiative of Heidelberg University; Marie Sklodowska-Curie Innovative Training Network Fellowship of the European Commission's Horizon 2020 Programme (No. 721559 AVA); European Research Council under the European Unions Seventh Framework Program FP7/2007-2013 (Grants Nos. 291242 and 277762); European Union's Horizon 2020 research and innovation programme under the Marie Sklodowska-Curie grant agreement ANGRAM No. 748826; Austrian Ministry for Science, Research, and Economy; Research Council of Norway; Bergen Research Foundation; John Templeton Foundation; Ministry of Education and Science of the Russian Federation and Russian Academy of Sciences and the European Social Fund within the framework of realizing the project, in support of intersectoral mobility and quality enhancement of research teams at Czech Technical University in Prague (Grant No. CZ.1.07/2.3.00/30.0034).

\bibliographystyle{apsrev}
\bibliography{aegis_biblio}

\end{document}